\documentclass[superscriptaddress,showpacs,aps,twocolumn,prb,floatfix]{revtex4}
\usepackage{amssymb}
\usepackage{amsmath}
\usepackage[dvips]{graphicx}
\usepackage{subfigure}
\usepackage{bm}
\usepackage{color}
\begin{document}

\title{Kondo physics in a Ni impurity embedded in O-doped Au chains}
\author{S. Di Napoli}
\affiliation{Departamento de F\'{\i}sica de la Materia Condensada, GIyA-CNEA, Avenida
General Paz 1499, (1650) San Mart\'{\i}n, Pcia. de Buenos Aires, Argentina}
\affiliation{Consejo Nacional de Investigaciones Cient\'{\i}ficas  y T\'ecnicas, CONICET, Buenos Aires, Argentina}

\author{M.A. Barral}
\affiliation{Departamento de F\'{\i}sica de la Materia Condensada, GIyA-CNEA, Avenida
General Paz 1499, (1650) San Mart\'{\i}n, Pcia. de Buenos Aires, Argentina}
\affiliation{Consejo Nacional de Investigaciones Cient\'{\i}ficas  y T\'ecnicas, CONICET, Buenos Aires, Argentina}

\author{P. Roura-Bas}
\affiliation{Departamento de F\'{\i}sica de la Materia Condensada, GIyA-CNEA, Avenida
General Paz 1499, (1650) San Mart\'{\i}n, Pcia. de Buenos Aires, Argentina}
\affiliation{Consejo Nacional de Investigaciones Cient\'{\i}ficas  y T\'ecnicas, CONICET, Buenos Aires, Argentina}

\author{L.O. Manuel}
\affiliation{Consejo Nacional de Investigaciones Cient\'{\i}ficas  y T\'ecnicas, CONICET, Buenos Aires, Argentina}
\affiliation{Instituto de F\'{\i}sica Rosario (CONICET-UNR), Rosario, Argentina}

\author{A.M. Llois}
\affiliation{Departamento de F\'{\i}sica de la Materia Condensada, GIyA-CNEA, Avenida
General Paz 1499, (1650) San Mart\'{\i}n, Pcia. de Buenos Aires, Argentina}
\affiliation{Consejo Nacional de Investigaciones Cient\'{\i}ficas  y T\'ecnicas, CONICET, Buenos Aires, Argentina}

\author{A.A. Aligia}
\affiliation{Centro At\'omico Bariloche and Instituto Balseiro, Comisi\'on Nacional de 
Energ\'{\i}a At\'omica, 8400 Bariloche, Argentina }

\begin{abstract}
By means of \textit{ab initio} calculations we study the effect of O-doping of Au chains containing 
a nanocontact represented by 
a Ni atom as a magnetic impurity. In contrast to pure Au chains, we find that with
a minimun O-doping the $5d_{xz,yz}$ states of Au are pushed up, crossing the Fermi level. We also find that 
for certain O configurations, the Ni atom has two holes in the degenerate  $3d_{xz,yz}$ orbitals, 
forming a spin $S=1$ due to a large Hund interaction. The coupling between the $5d_{xz,yz}$ Au bands and the
$3d_{xz,yz}$ of Ni states leads to a possible realization of a two-channel $S=1$ Kondo effect. While 
this kind of Kondo effect is commonly found in bulk systems, it is rarely observed in low dimensions. The 
estimated Kondo scale of the system lies within the present achievable experimental resolution in 
transport measurements. 
Another possible scenario for certain atomic configurations is that one of the holes resides 
in a $3d_{z^2}$ orbital, leading to a two-stage Kondo effect, the second one with SU(4) symmetry.
\end{abstract}

\maketitle

\section{Introduction}

In modern nanoscience, tailoring the electronic transport through atomic-size conductors has turned into a duty, 
as it is a powerful tool for detecting nanomagnetism. Atomic-size contacts can be experimentally 
obtained by different techniques, particularly in a mechanically controllable break junction (MCBJ) experiment, 
where the formation of one dimensional atomic chains of several kinds of elements is possible.~\cite{Ruitenbeek98,
Ohnishi98,Ruitenbeek01,Rubio01,Ryu06} Since the achievement of the first free-standing atomic chains of gold atoms 
in 1998,~\cite{Ruitenbeek01,Ohnishi98} the search for other elements that could also form atomic chains has been 
intense and active. For instance, in the last few years it has been possible to strengthen the bonds in a 
suspended chain and to achieve a higher probability of chain producibility by adding external absorbates during 
the chain formation process. It is known that low-coordinated atoms are chemically more reactive than in 
bulk,~\cite{Barnett04} thus, chains are expected to be even more reactive than nanoparticles, opening the 
possibility for molecular absorbates to dissociate, even at low temperatures. For instance, oxygen (O) atoms are 
expected to be incorporated in the chains, as predicted in several previous works.~\cite{Novaes06,Bahn02,Thijssen06,
Dinapoli12,Aradhya13,Cespedes15}

Due to outstanding experimental achievements in the last decades, it is nowadays possible to design nanodevices as 
tools for detecting nanomagnetism by conductance measurements through atomic metal contacts. One can indirectly 
sense the presence of magnetism by detecting zero-bias anomalies, usually originated in the Kondo screening of the
spin, if a magnetic impurity is bridging the contact. 

In previous works, Lucignano \textit{et al.}~\cite{Tosatti09} and Miura \textit{et al.}~\cite{Tosatti08} studied the 
electronic structure and the Kondo conductance through a Ni impurity embedded in a monoatomic Au wire. The Ni atom in
the Au chain has two low-energy geometries: bridge (B) and substitutional (SUB). While in the B configuration, an
usual one-channel Kondo (1CK) effect of spin $S=1/2$ was reported theoretically, no Kondo physics was predicted in the 
SUB  configuration.
This is due to the fact that in the SUB geometry (of higher total energy but probably accessible at large stress,
as it is the case in a MCBJ experiment), the empty spin-down Ni state
orbitals are a $3d_{xz,yz}$ degenerate pair with angular momentum projection $\left|m\right|=1$. In this 
geometry the Ni impurity has $S\simeq 1$ and the magnetic orbitals are 'spectators' as they cannot be screened due to 
their orthogonality to the $m=0$ Au conduction band. Indeed, transport experiments in Au chains indicate that the 
conduction channel is a single $6s$ band.~\cite{Rego03, Ruitenbeek03}

As it was previously mentioned, O impurities in the Au chain are expected during its real formation under an open atmosphere. 
A dramatic effect of the incorporated O impurities to the Au chains was found in our recent
{\it ab initio} calculations.~\cite{Dinapoli13} They can modify the 
band structure of the metallic host, pushing up the bands that are close to the Fermi level, 
thus establishing conduction through the $5d_{xz}$  and $5d_{yz}$ electrons of Au. 
In these conditions, the spin of the Ni impurity, in the SUB geometry, formed by the localized $3d_{xz,yz}$ electrons, 
can be screened by conduction electrons of the O-doped Au chain and Kondo physics is expected.
A similar feature was obtained in O-doped Au chains containing a magnetic $S=3/2$ Co 
impurity.~\cite{DinapoliPRL,DinapoliPRB} 
Remarkably, the examples of Ni-Au-O and Co-Au-O systems are expected to be realizations of Kondo models 
where an arbitrary spin $S$ is screened by $n$ channels, and were solved exactly in the 90's by the Bethe 
ansatz~\cite{Schlottmann93}. 
For $n=2S$ the models present Fermi liquids properties, while for $n>2S$ non-Fermi-liquid behavior is obtained 
and for $n<2S$ the systems are singular Fermi liquids.\cite{mehta}
The presence of anisotropy render models not solvable by the Bethe ansatz and modify
the low-temperature behavior.~\cite{DinapoliPRL,DinapoliPRB,corna}

In this paper, we focus on a Ni magnetic impurity embedded, in a SUB configuration, within an O-doped Au chain. We 
study the effects of the 
presence of O impurities on the symmetry of the Au conduction bands  close to the Fermi level as well as on
the spin state of the Ni impurity. 
This symmetry determines the nature of the screening of  the impurity spin, giving rise to the possibility of 
the system to exhibit some kind of Kondo effect. 
In particular, we found a stable configuration in which the model that describes the 
system corresponds to a spin $S = 1$ screened by two conduction channels. 
Using parameters coming from \textit{ab initio} calculations, we build the
model Hamiltonian which corresponds to a generalized Anderson impurity model. 
After a Schrieffer-Wolff transformation we prove that the
effective low energy model is a $S=1$, 2-channel Kondo model. We analyze this model at zero temperature 
and give an estimation of the expected Kondo temperature ($T_K$) for this system.

For other atomic configurations, corresponding to metastable states, the \textit{ab initio} results suggest
a six-fold degenerate ground state of the Ni atom, with one hole in a $3d_{z^2}$ orbital and another 
one in either a $3d_{xz}$ or a $3d_{yz}$ orbital in an $S=1$ state. This is analogous to the 
situation of Fe iron(II) phtalocyanine molecules on Au(111), where a two-stage Kondo effect
takes place, screening first the spin $1/2$ of the $3d_{z^2}$ hole and then that of the 
orbitally degenerate remaining hole in an SU(4) Kondo effect.~\cite{mina}

The paper is organized as follows. 
In section \ref{abinitio} we provide the details of our DFT first-principles calculations
and show the band structure for the different  hosts and the consequences of these structures on the systems 
with an embedded Ni impurity. In section \ref{model} we present the Kondo model, obtained by a Schrieffer-Wolff
transformation of the $S=1$ two-channel Anderson model, and obtain the Kondo temperature.
Finally, a summary and discussion are given in section \ref{conclusions}.

\section{\textit{Ab initio} results: symmetry of the A\lowercase{u} conduction bands and N\lowercase{i} holes}
\label{abinitio}

We perform \textit{ab initio} calculations based on density functional theory (DFT) using the full potential 
linearized augmented plane waves method, as implemented in the WIEN2K code.~\cite{Wien} 
The generalized gradient approximation for the exchange and correlation potential in the parametrization of 
PBE~\cite{PBE96} and 
the augmented plane waves local orbital basis are used. The cutoff parameter which gives the 
number of plane waves in the interstitial region  is taken as $R_{mt}*K_{max} = 7$, where $K_{max}$ is the value of the 
largest reciprocal lattice vector used in the plane waves expansion and $R_{mt}$ is the smallest muffin tin radius 
used. The number of \textbf{k} 
points in the Brillouin zone is enough, in each case, to obtain the desired energy  and charge precisions, namely 10$^{-4}$ Ry and 
10$^{-4}$e, respectively. 
The muffin-tin radii were set to $2.23$~bohr for Ni, $1.91$~bohr for Au atoms and $1.69$~bohr for the O 
impurities. In all the studied cases we consider a  periodically repeated hexagonal lattice with $a=b=15$~bohr and  the coordinate 
system is fixed in such a way that the chain axis is aligned with  $z$. The $a$ and $b$ distances in the supercell were 
checked to be large enough
to avoid artificial interactions between the periodic replicas of the wire.\\

\begin{figure}
\begin{center}
\includegraphics[width=1.\columnwidth]{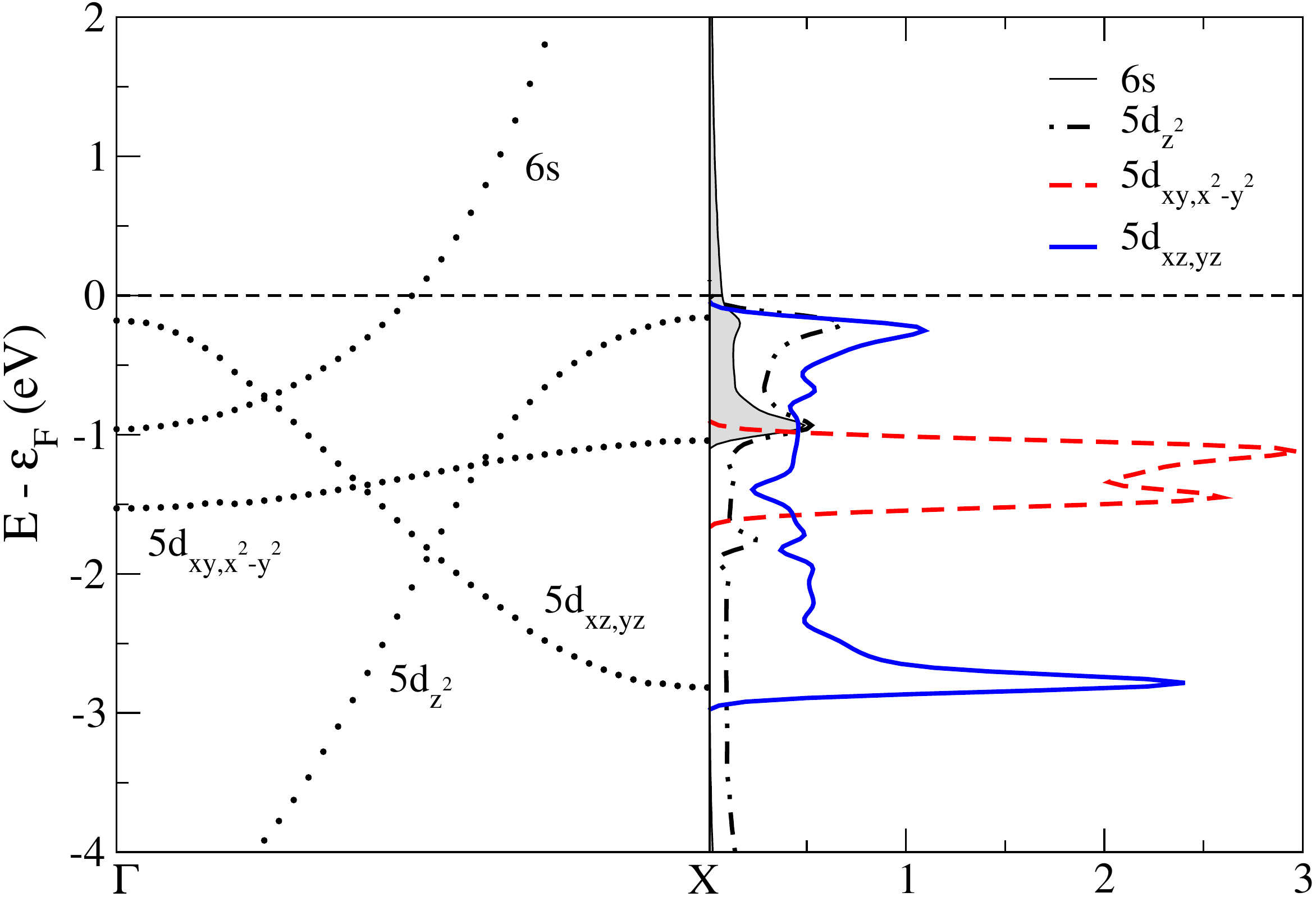} 
\end{center}
\caption{(left) Electronic band structure and (right) orbital decompositioned projected density of states, PDOS, 
of the Au monoatomic chain at the GGA equilibrium distance $d_{Au-Au}^{eq}=4.9285$ bohr. The Hubbard U parameter set 
in this calculations is $U=4eV$.}
\label{Au-bands}
\end{figure}

As a first step, we analyze the case of a Au monoatomic chain and its band structure. Usually, DFT calculations yield a spurious 
magnetization in  Au wires, due to self-interaction errors, which shift the $5d_{xz,yz}$ ($\left|m\right|=1$) bands of the Au
wire up to the Fermi energy. One route to avoid this
spurious result is to increase the Au-Au distance in the chains during the calculations.~\cite{Tosatti08, Tosatti09} 
We pick up another route to tackle this problem and, as suggested by Sclauzero and Dal Corso,~\cite{DalCorso13} 
we include a Hubbard U correction of $U=4eV$ in the $5d$-electron Au manifold. As it can be seen in 
Fig.~\ref{Au-bands}, the mentioned magnetization is completely suppressed in the 
chain at the equilibrium distance of $d_{Au-Au}^{eq}=4.9285$~bohr. \\  

\begin{figure}
  \centering
  \includegraphics[width=0.8\columnwidth]{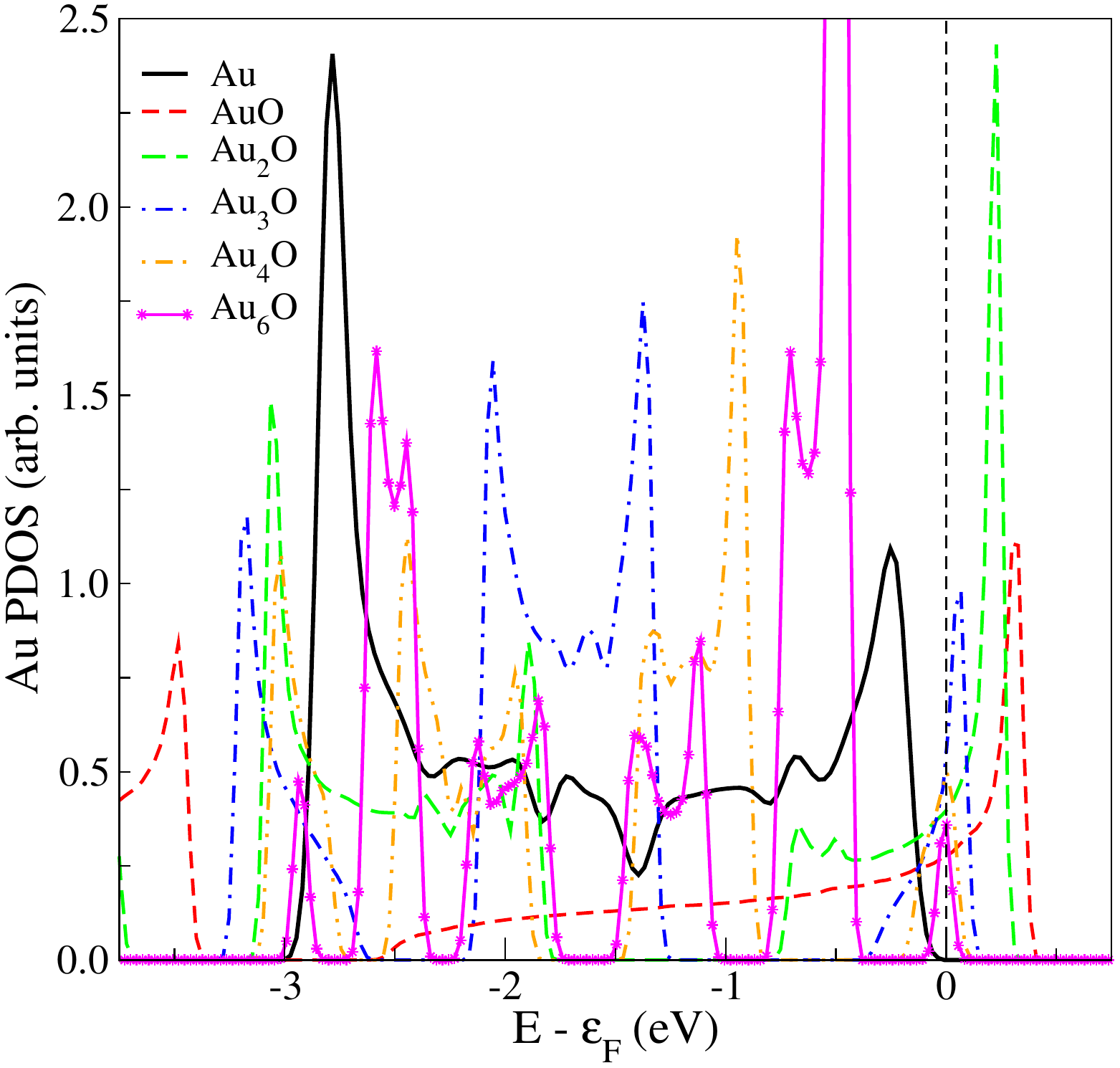} \includegraphics[width=1.\columnwidth]{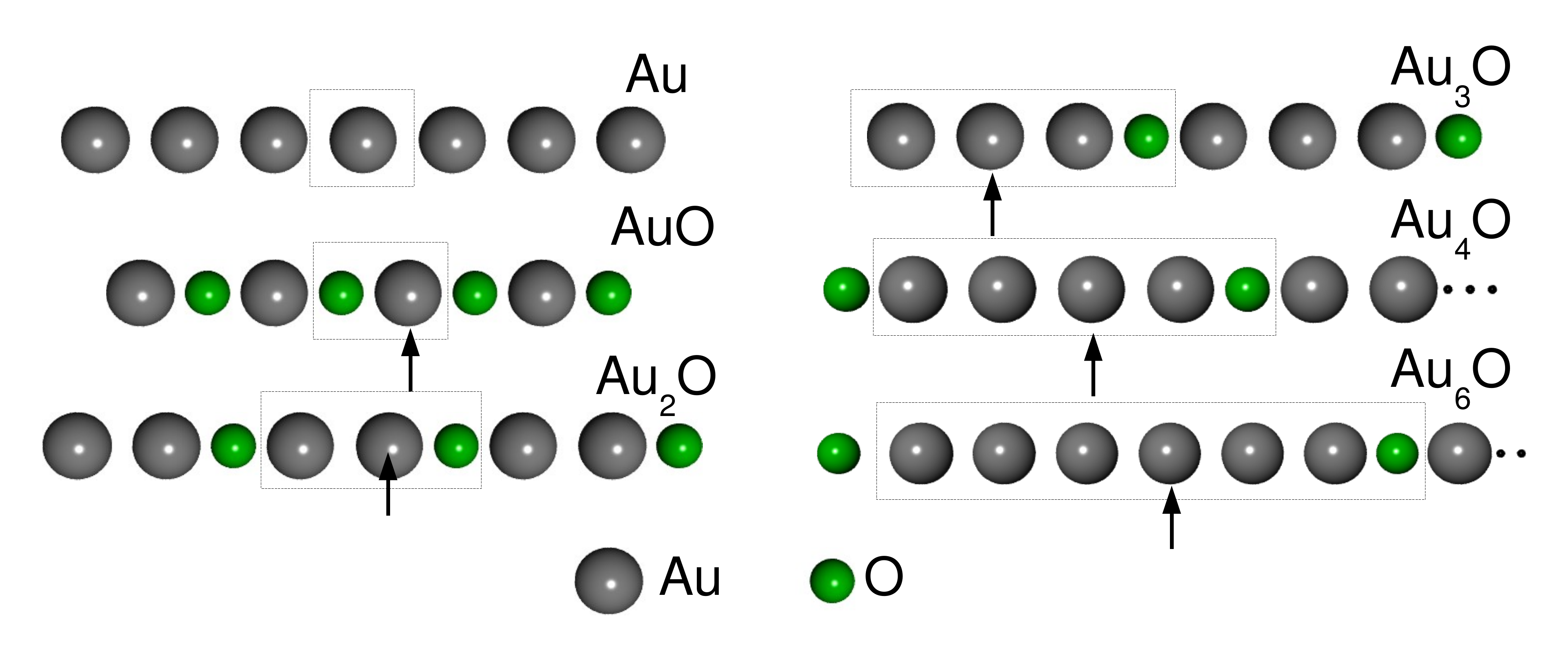}  
  \caption{(Color online) (Top) $5d_{xz,yz}$ projected density of states of the Au atom located farthest away from the 
O atom. (Bottom) Schematic representation of the chains. The unit cell is shown for each case and the arrows indicate 
the Au atom whose projected density of states is shown in the top panel.}
\label{AuO-dos}
\end{figure}

In O-doped Au chains, we relax the Au-O distance for the case of AuO diatomic chain (two-atom unit cell) 
and, afterwards,  we take the same bond length, 
$ d_{Au-O}^ {eq}=3.625$~bohr, for all the studied chains. We consider several  O-dopings to find which the 
minimal amount of O is needed to push the projected $5d_{xz,yz}$ density of states of all Au atoms towards the Fermi level. 
In this way, the $\left|m\right|=1$~-symmetry conduction channel through all the Au atoms in the chain is opened. 
In all the studied cases - i.e.: 50\% (AuO), 33.3\% (Au$_2$O), 25\% (Au$_4$O) and 14\% (Au$_6$O) - we include the same Hubbard $U=4eV$
in the $5d$-electrons of Au and perform self-consistent calculations. We find that the $5d_{xz,yz}$ orbitals of Au, up 
to the third neighbors of the O impurity, cross the Fermi level, due to the large hybridization with the oxygen $2p_{x,y}$ states.
This is shown in Fig.~\ref{AuO-dos}, where we plot the $5d_{xz,yz}$ density of states of the Au atom located 
farthest away from the O atoms, for some selected dopings. From this we can conclude that an O-doping of at least 14\% is necessary 
to transform the Au bands into conducting channels. We consider below   a doping  of $\approx$ 19 \%  (3/16).

\begin{table*}[ht!]
\caption{\label{occ} Symmetry-dependent $d-$band minority spin fillings of the Ni atoms (in electrons) for the 
  selected studied cases. The color coding of the schematic representation of the chains is the one presented
  in Fig.~\ref{PDOS}(a).}
\begin{ruledtabular}
\begin{tabular*}{\hsize}{l@{\extracolsep{0ptplus1fil}}ccccc}
Case	&Unit cell's schema &$n(3d_{z^2})$& $n(3d_{xy,x^2-y^2})$ &  $n(3d_{xz,yz})$ & $\mu_{Ni} (\mu_B)$\\
\colrule
\textit{sym O-4$^{th}$-nn}&     \includegraphics[width=0.5\columnwidth]{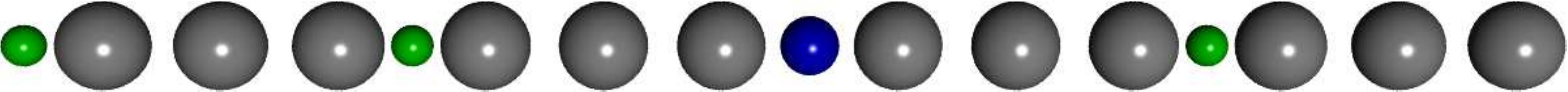}  & 0.92 & 1.87 & 0.56 & 1.37\\
\textit{sym O-3$^{th}$-nn}&     \includegraphics[width=0.5\columnwidth]{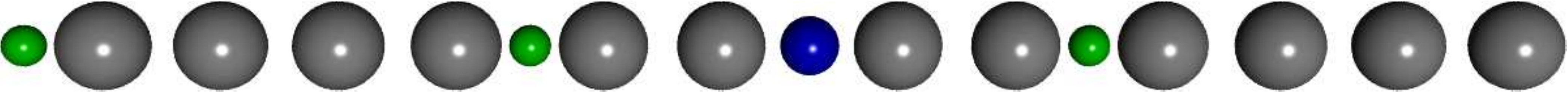}   & 0.86 & 1.69 & 1.00 & 1.11\\
\textit{sym O-2$^{th}$-nn}&     \includegraphics[width=0.5\columnwidth]{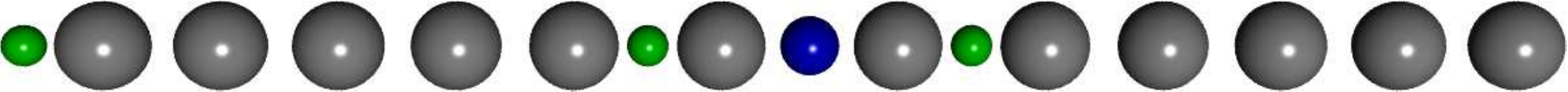}   & 0.63 & 1.60 & 1.21 & 1.33\\
\textit{no-sym 1O-2$^{th}$-nn}& \includegraphics[width=0.5\columnwidth]{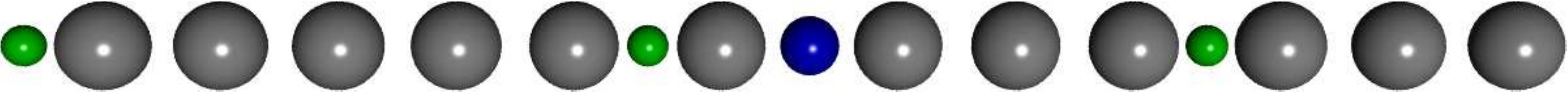} & 0.67 & 1.58 & 1.18 & 1.31\\
\textit{no-sym 1O-3$^{th}$-nn}& \includegraphics[width=0.5\columnwidth]{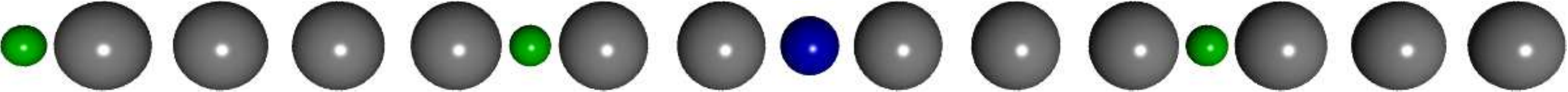} & 0.91 & 1.82 & 0.72 & 1.26\\
idem above metastable& \includegraphics[width=0.5\columnwidth]{NiAu12O3_nosym3nn} & 0.27 & 1.53 & 1.67 & 1.26 \\
\end{tabular*}
\end{ruledtabular}
\end{table*}

Once the minimum amount of O present in the Au chains is set, we include the Ni impurity in our calculations. 
The question that arises then is,  where to put the O atoms and the Ni impurity within the Au chain. We assume that
the O atoms are not bonded to the Ni impurity and relax the Au-Ni bond length for a diatomic Ni-Au chain, 
giving $d_{Au-Ni}^{eq}= 4.4805$~bohr, in excellent agreement, within a $0.05\%$, with that obtained by 
Miura \textit{et al.}.~\cite{Tosatti08}.
For a chain containig 12 Au atoms, 3 O atoms and 1 Ni impurity within the unit cell, we take
the corresponding bond lengths as $d_{Au-Ni}^{eq}= 4.4805$~bohr, $ d_{Au-O}^ {eq}=3.625$~bohr and  $d_{Au-Au}^{eq}=4.9285$~bohr.
In consequence, the Ni-Ni distance in the $z$-direction is larger than $70$~bohr and, therefore, we assume that the Ni-Ni magnetic 
interaction is negligible.\\  
To find out if the O's position, with respect to the Ni atom, changes the spin-state of the Ni impurity or
its hole's symmetries, we explore several configurations, always keeping the same amount of O in the Au chain.
In table~\ref{occ} we introduce the different cases studied and  their corresponding Ni magnetic moments 
as well as the occupation numbers (in electrons) obtained by integrating   the 
minority-band  separated in the different symmetries, within the Ni-muffin-tin sphere. 
The last entry corresponds to a metastable solution, whose energy is 85 meV higher than that of the previous entry.
Note that for the $3d_z^2$-symmetry the maximal occupation 
could be as much as $n(3d_z^2)=1$, whereas $n(3d_{xy,x^2-y^2})= n(3d_{xz,yz}) \le 2$  as they both are 2-fold 
degenerate states.
Based on the data shown in the table, we can distinguish between two extreme cases for the stable solutions: the one 
called \textit{sym O-4$^{th}$-nn} and the
denominated \textit{sym O-2$^{th}$-nn}. In the former case, the Ni atom is placed in the center of the O-doped Au chain and it
has two O atoms symmetrically located as fourth neighbors. The $3d_{z^2}$ orbital is almost completely filled and the 
two-fold degenerate orbitals $3d_{xz,yz}$ are almost empty. As the O atoms approach the Ni impurity, 
the $3d_{xz,yz}$ orbitals increase their filling while the $3d_{z^2}$ orbital moves towards the Fermi level decreasing 
its occupancy until it becomes almost half-filled in the case called \textit{sym O-2$^{th}$-nn}. In this last case, the 
Ni impurity has two O atoms at second neighbor positions.
Note that for the metastable configuration, with one O atom as third and another as forth nearest-neighbors 
(\textit{no-sym 1O-3$^{th}$-nn}), the minority $3d_{z^2}$ state is almost empty. The small energy difference (85 meV) 
between the metastable and stable configurations suggests a near degeneracy between the triplet 
of the Ni 3d$^8$ configuration with two holes occupying the $m= \pm 1$ states (or $3d_{xz,yz}$ orbitals) 
and the two triplets  with one hole in the $m=0$ ($3d_{z^2}$) state and another one with $m= \pm 1$. In the calculations there 
is always a finite splitting, larger than 0.5 eV, positive or negative of the peaks in the 
minority spectral density for $m=0$ and $m= \pm 1$, but this is probably due to the exchange and correlation potential 
included in the DFT, since a partial occupation of one state pushes the other up in energy.

\begin{figure}
 \centering
 \subfigure[Case \textit{sym O-4$^{th}$-nn}: Two O atoms are symmetrically positioned as Ni-fourth neighbors.]
 {
 \begin{tabular}{c}
 \includegraphics[width=0.8\columnwidth]{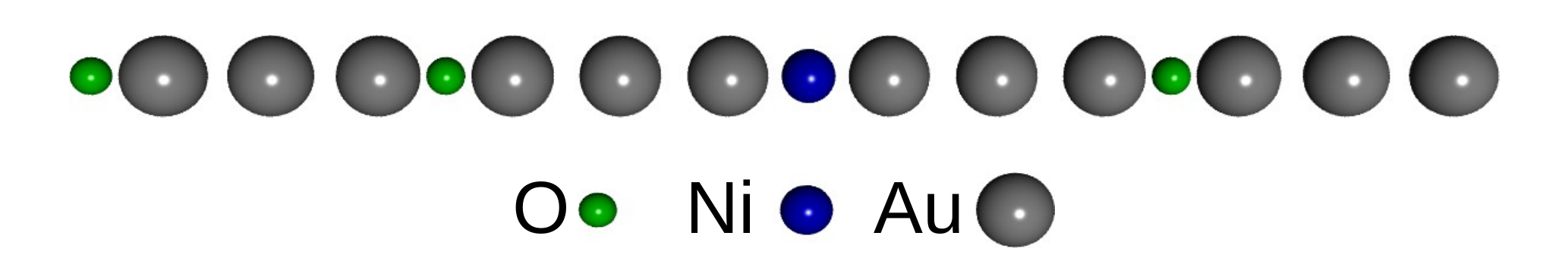}\\
  \includegraphics[width=0.7\columnwidth]{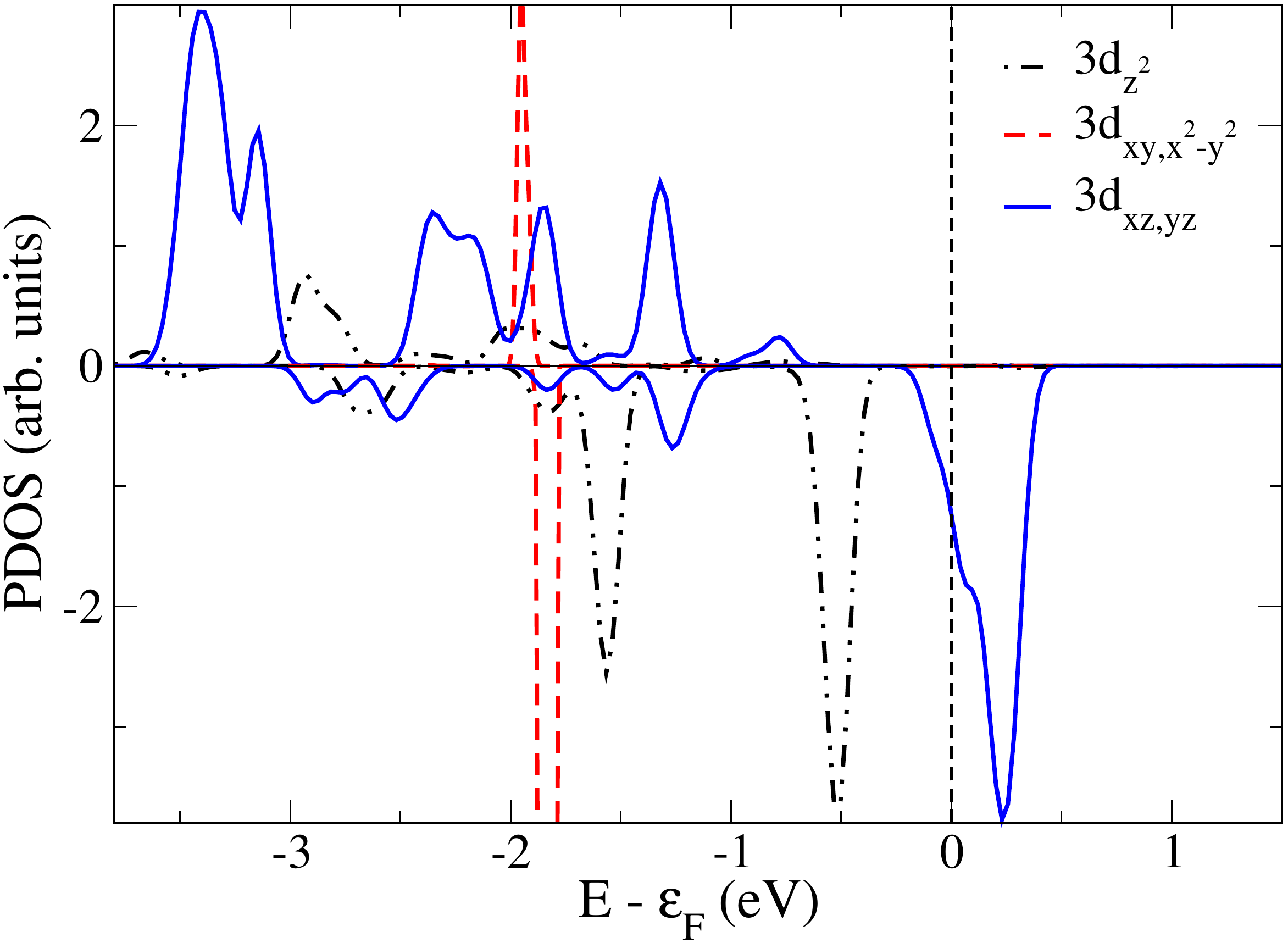}
 \end{tabular}
  }
  \\
  \subfigure[Case \textit{sym 1O-2$^{th}$-nn}: Two O atoms as Ni-second neighbors.]
  {
   \begin{tabular}{c} 
   \includegraphics[width=0.7\columnwidth]{NiAu12O3_sym2nn}\\
   \includegraphics[width=0.7\columnwidth]{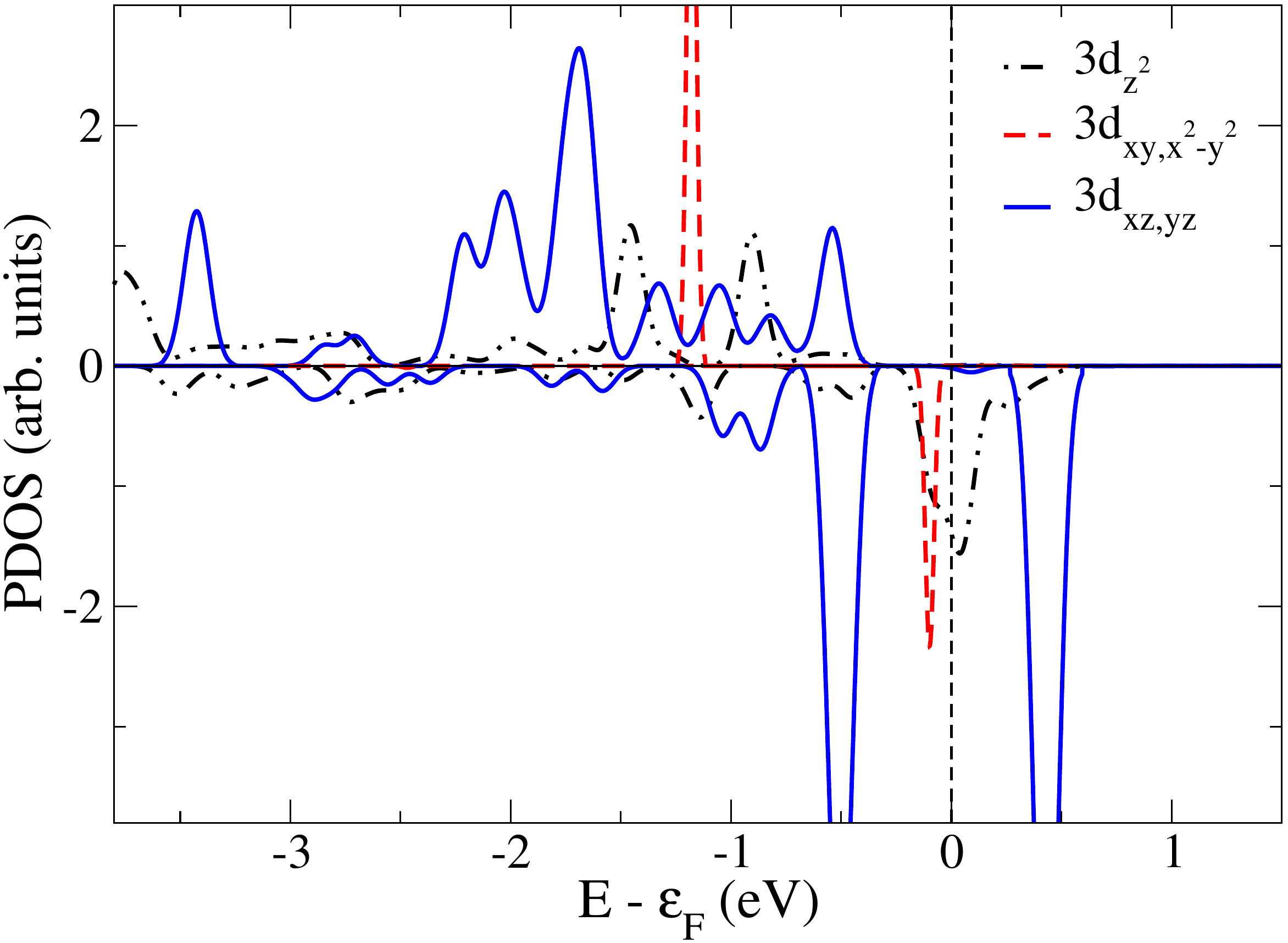}
 \end{tabular}
  }
  \caption{Ni impurity in a 19\% O-doped Au chain. In each case, the unit cell and the density of states projected onto 
    different symmetries are shown.}
\label{PDOS}
\end{figure}

In Fig.~\ref{PDOS} we show the partial density of states projected onto the different symmetries of the Ni
impurity, for two of the above mentioned cases. The Ni impurity develops a magnetic moment of $\mu_{Ni}=1.37\mu_B$ in 
the \textit{sym O-4$^{th}$-nn} configuration and, as also found for the Ni impurity embedded in a Au 
monowire,~\cite{Tosatti08}, the Au atoms at both sides of the Ni atom develop a small induced magnetic moment of around 
$0.18\mu_B$. 
The calculated Ni magnetic moment is  $\mu_{Ni}=1.33\mu_B$ in the \textit{sym O-2$^{th}$-nn} case, and the induced magnetic
moment in the Au atoms located at both sides of the Ni impurity is around $0.31\mu_B$. The enhancement of this last induced 
magnetic moment is due to the proximity of the O atoms (See sketch of Fig.~\ref{PDOS}(b)).\\
In spite of the fact that the spin state of the Ni impurity in both configurations is almost the same, the hole's 
symmetries
are different. It can be seen from Fig.~\ref{PDOS}(a) that the empty spin-down Ni orbitals (holes) are the $3d_{xz,yz}$, 
while the other orbitals remain occupied. The different band-fillings change, depending on the position of the O impurities. 
Upon approaching the O atoms towards the Ni impurity, the $3d_{xz,yz}$ orbitals begin to be filled, 
while the $3d_{z^2}$ one begins to be unoccupied, as shown in Fig.~\ref{PDOS}(b).
One also observes in this figure that the minority $3d_{z^2}$ state lies at the Fermi level and the density 
corresponding to the $3d_{xz,yz}$ states is split in two around the Fermi level, suggesting a near degeneracy of 
the $m=0$ and $m= \pm 1$ states as in the case when  the O atoms are third and forth nearest-neighbors of Ni,  already discussed above.
In all the treated  cases, the $3d_{xy,x^2-y^2}$ states are the most localized, as they lie 
perpendicular to the chain, thus having a small hybridization. Nevertheless, we observe that the position of these filled
degenerate orbitals might change depending on O proximity.

In order to determine the anisotropy constant $D$, taken into account in the term $H_D=D M_2^2$ of the  
Hamiltonian defined in Eq.~\ref{ham} of Section~\ref{model}, we introduce the spin orbit interaction in the Ni impurity 
and perform self consistent calculations within the fully relativistic approximation. For the \textit{sym O-4$^{th}$-nn} 
case we obtain a magnetocrystalline anisotropy energy, MAE, equal to $E_{\parallel} - E_{\bot} \sim 10$ meV, where 
$E_{\parallel}$ ($E_{\bot}$) is the total energy corresponding to the total magnetization pointing along (perpendicular 
to) the chain's axis. Therefore, we obtain that the easy magnetization axis is perpendicular to 
the chain. To calculate the value of $D$ we proceed as in Ref. \onlinecite{barral}. The expectation value
of the anisotropy in the state $|11 \rangle$ of maximum projection in the $z$ direction is clearly
$\langle 11|H_D|11\rangle=D$. By rotating this state, one obtains that the state of maximum projection
in the $x$ direction is $|11x \rangle= |10 \rangle/ \sqrt{2}+(|11 \rangle +|1-1 \rangle)/2$, and
$\langle 11x|H_D|11x\rangle=D/2$. From the energy difference $D\sim 20$ meV.

\section{Effects of correlations inside the N\lowercase{i}}
\label{corr}

In this section we investigate the effects of correlations in an isolated 3d$^8$ configuration
on the relative stability of the two possible ground states discussed above: spin triplet 
with total angular momentum projection $L_z=0$ with one hole with $m=1$ and the other with
$m=-1$, or two triplets with $L_z= \pm 1$ (one hole with $m=0$ and the other with
$m= \pm 1$). We also provide an alternative estimation of the anisotropy parameter $D$.
Since the GGA underestimates correlations that affect the orbital polarization of the 
$d$ states,\cite{eschrig, erik,nico,sd3}
this calculation is an important complement to the {\it ab initio} results presented above.

We have solved exactly the $45 \times 45$ matrix of the Hamiltonian corresponding to the $3d^8$ configuration,
containing crystal-field splitting and all correlations inside the $d$ shell, as described 
for example in Refs. \onlinecite{sd3,kroll}. 
In a later step, to calculate $D$, we include the spin-orbit coupling
$\hat{H}_\mathrm{SOC} =
\lambda \sum_{i}\hat{\mathbf{l}}_{i} \cdot \hat{\mathbf{s}}_{i}$.
We have considered the Coulomb integrals $F_{2}=0.16$ eV, $F_{4}=0.011$ eV, and $\lambda =0.08$ eV
from a fit of the
low energy spectra of late transition metal atoms.

For the crystal-field splitting, we took two different sets of parameters.
From the {\it ab initio} results presented in Fig. 3 (a) for the minority states, we estimate 
-1.85, -1 and -0.3 eV for the on-site energy of the states with $m= \pm 2$, 0 and $\pm 1$
respectively. For the latter, which is unoccupied we have subtracted
a correlation term included in the GGA, which we estimate to be 0.5 eV.
In the second set we assume that the $m=0$ and $m= \pm 1$ states are degenerate, with the same 
average distance to the $m= \pm 2$ states as before. This is equivalent to take
-1.2 eV for the energy of the electrons with $x^2-y^2$ and $xy$ symmetry and 0 for the remaining 
three. Note that in the last case, in the absence of interactions, the above mentioned
two-hole possible ground states with $L_z=0$ and $L_z= \pm 1$ are degenerate.
We find that interactions stabilize the triplet with $L_z=0$ by 0.30 eV.
For the first set of parameters, the stabilization energy for this triplet, 
which is 0.7 eV in the absence of interactions, is increased to 0.88 eV in
the present calculation. The difference 0.18 eV can be ascribed to the effect of correlations.

When spin-orbit is included we obtain $D=8.5$ (16.7) meV for the first (second) set of parameters.
The sign and order of magnitude agree with the {\it ab initio} results.

\section{N\lowercase{i} impurity within an O-doped A\lowercase{u} chain: Kondo effect}
\label{model}

According to the results of the previous section, different model Hamiltonians 
can be formulated depending on the O position relative to the Ni impurity.
In general, for the Ni $3d^8$ configuration, the triplet ground state with $L_z=0$
is the most stable in the {\it ab initio} calculations and is further stabilized by 
correlations. Therefore in this section, we analyze this case in more detail.
The situation with two degenerate triplets with $L_z= \pm 1$ is very similar to
the physics of an organic molecule in a metallic substrate and will be addressed to
in Section \ref{conclusions}

\begin{figure}[ht!]
\centering
\includegraphics[width=0.9\columnwidth]{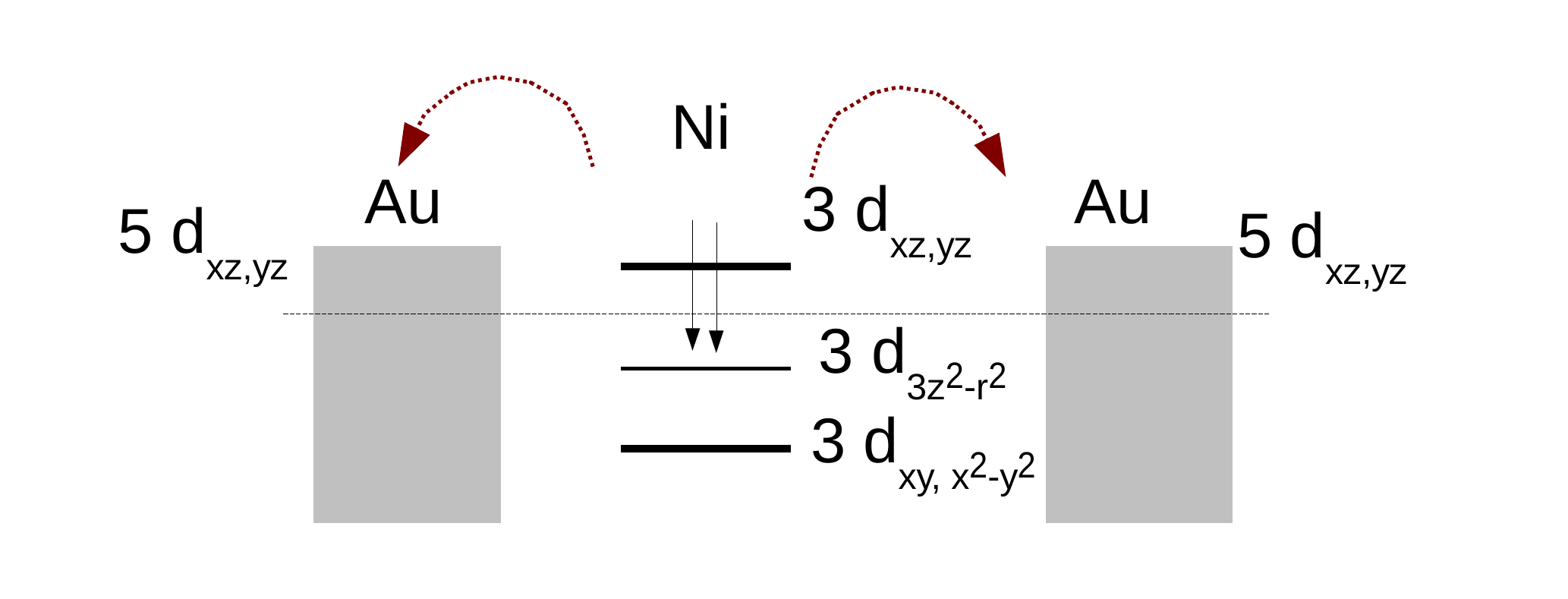}
\caption{Schematic representation of Ni-holes and their tunneling to the Au-leads for the case \textit{sym O-4$^{th}$-nn}.}
\label{esq_O4nnsym}
\end{figure}

To be specific, the system we are interested in, represented by the first row of table~\ref{occ}, namely \textit{sym 
O-4$^{th}$-nn}, is the one in which the symmetries of the Ni holes are well defined in the orbitals $\alpha=yz, xz$. 
Moreover, this case presents a larger hybridization with the host as it can be seen in the broader density of states of 
these orbitals, see Fig.~\ref{PDOS}, leading to a larger Kondo scale.    
Taking into account the \textit{ab initio} results presented in the previous section, 
the charge fluctuations in the Ni impurity are mainly due to the interchange between 
the state with two holes and virtual excitations to states with only one hole in 
either the $3d_{yz}$ or $3d_{xz}$ orbital (See Fig.~\ref{esq_O4nnsym}). Furthermore, both configurations 
are magnetic with spin $S=1$ and $S=1/2$ for two and one holes respectively.     
The spin-orbit coupling (SOC) in the Ni atom induces a splitting $D$ between the 
projections $S_z = ±1$ and $S_z = 0$ of the triplet that belongs to the total spin $S=1$.

With these information, we propose a Hamiltonian that describes the system, and that is given 
by~\cite{aligia-balseiro-proeto}

\begin{eqnarray} \label{ham}
H&=&\sum_{M_{2}}(E_{2}+
D M_{2}^{2})|M_{2}\rangle \langle
M_{2}|+\sum_{\alpha M_{1}}E_{\alpha}|\alpha M_{1}\rangle \langle \alpha M_{1}| + \nonumber\\
&&\sum_{\nu k\alpha \sigma }\epsilon _{\nu k}c_{\nu k\alpha \sigma
}^{\dagger }c_{\nu k\alpha \sigma }+\\
&&\sum_{M_{1}M_{2}}\sum_{\alpha\nu k\sigma }V_{\nu }\langle 1
M_{2}  |\frac{1}{2}\frac{1}{2}M_{1}\sigma \rangle 
( | M_{2}\rangle \langle \alpha M_{1}|c_{\nu k\alpha \sigma }+\mathrm{H.c.} ),  \nonumber 
\end{eqnarray}

where $E_i$ and  $M_i$ represent the energies and the spin projections along the chain, chosen as the 
quantization axis, of states with $i=1, 2$ holes in the $3d$ shell of the Ni impurity. The state with two holes
and maximum spin projection is denoted by $\vert 1 \rangle = 
\hat{d}_{xz\downarrow}^{\dagger}\hat{d}_{yz\downarrow}^{\dagger}
\vert 0 \rangle$, where $\vert 0 \rangle$ represents the $3d^{10}$ configuration and the operator
$\hat{d}_{\alpha\sigma}^{\dagger}$ creates a hole with symmetry $\alpha=xz, yz$ and spin 
projection $\sigma$.
The states with one hole in the Ni atom can be constructed by removing an $\alpha$ hole. 
The other relevant states with two and one holes can be obtained by using the spin lowering operators.

The operator $c_{\nu k\alpha \sigma}^{\dagger }$ creates a hole in the $5d$ shell of the Au atom with symmetry 
$\alpha$, where $\nu=L, R$ denotes the left or the right side of the Ni atom, respectively.  The hopping $V_\nu$ 
characterizes the tunneling between the Ni and Au states.
\vspace{1.0cm}

\textit{The Kondo regime}.

To describe the spin fluctuations, ie, the Kondo regime in which 
the charge fluctuations are frozen, we perform a Schrieffer-Wolff transformation. The effective 
low energies' Hamiltonian is then

\begin{eqnarray} \label{ham_eff}
H_{eff}&=&\sum_{\nu \alpha\nu}\epsilon _{\nu k}c_{\nu k\alpha \sigma
}^{\dagger }c_{\nu k\alpha \sigma }+
\nonumber \\
&& \sum_{\alpha\nu}\frac{\vert V_\nu \vert ^2}{2}\{ \frac{|1\rangle \langle 1|+|-1\rangle \langle -1|}
{E_1 - D - E_2}+
\frac{|0\rangle \langle 0|}{E_1 - E_2} \}\hat{n}_{c\alpha\nu} + \nonumber \\
&&J_{\parallel}  \sum_{\alpha}S_z s_{\alpha}^{z} + J_{\bot}\sum_{\alpha} (S_x s_{\alpha}^{x} + S_y s_{\alpha}^{y} )
\end{eqnarray}

where $\hat{n}_{c\alpha\nu}=\sum_{k \sigma}c_{\nu k\alpha \sigma}^{\dagger }c_{\nu k\alpha \sigma }$
and $J_{\parallel}=\frac{\vert V \vert ^2}{E_1 - E_2 - D}$, 
$J_{\bot}=1/2(\frac{\vert V \vert ^2}{E_1 - E_2 - D}+\frac{\vert V \vert ^2}{E_1 - E_2})$ 
and $\vert V \vert^2 = ( \vert V_L \vert^2 + \vert V_R \vert^2)/2$.
The middle term represents a renormalization of the local energies and the potential scattering 
and are unimportant for understanding the Kondo effect. The final term represents the exchange interaction
between the impurity spin $S$ and the corresponding ones of the O-doped Au bands $s_{\alpha}$, being 
$J_{\parallel}$ and $J_{\bot}$ the longitudinal and perpendicular couplings. This exchange term corresponds
to the anisotropic Kondo Hamiltonian, $H_K$

\begin{eqnarray} \label{ham_K}
H_{K}&=&J_{\parallel}  \sum_{\alpha}S_z s_{\alpha}^{z} + 
J_{\bot}\sum_{\alpha} (S_x s_{\alpha}^{x} + S_y s_{\alpha}^{y} )
\end{eqnarray}

Notice that in the absence of any anisotropy interaction, $D=0$, the model 
reduces to $J~S \cdot (s_{yz}+s_{xz})$, which is a spin-1 two-channel, $n=2$, 
Kondo model with $J=\frac{\vert V \vert ^2}{E_1 - E_2}$, which is a Fermi liquid.~\cite{Schlottmann93,nozieres-blandin} 
 
The Kondo temperature for our model can be estimated from the
usual expression $T_K\sim e^{1/2J\rho_c(E_F)}$ in which $\rho_c(E_F)$ corresponds to
the value of the density of states of the conduction band at the Fermi level, $E_F$.
The only change with those associated to the spin-1/2 one-channel model enters in the
change of the exchange coupling $J$, a half in our case as compared to the spin-1/2 
one-channel.~\cite{s-1-coleman}

For a numerical evaluation in terms of the \textit{ab initio} parameters we use the 
fact that $J\rho_c(E_F)=\frac{\Delta}{\pi\epsilon_d}$, being $\Delta=\pi V^2 \rho_c(E_F)$ 
and $\epsilon_d=E_1-E_2\sim-E_{xz}$. We estimate $\Delta$ from the half maximum width
of the $3d_{\alpha}$ density of states, which corresponds to $2\Delta\sim0.23$ e$V$, 
and $\vert E_{xz}\vert\sim0.23$ e$V$ from its energy position. 
Thus, we estimate a Kondo temperature close to $T_K\sim400K$, and therefore the zero-bias anomaly should 
be observed in transport measurements.

As stated in Section~\ref{abinitio}, we find that the value of the anisotropy constant $D$ is of the order of
$D\sim 15$ me$V \sim 180 K$. 

Studies of similar models by the numerical renormalization group suggest that for $D < T_K$ 
(as in this case) an energy scale $D^*$ much smaller than $D$ is dynamically generated.
For example, for the underscreened $S=1$ Kondo model with positive $D$, 
$D^* \approx \exp[-c(T_K/D)^{1/2}]$. While for $D=0$ the conductance reaches its maximum value at $T=0$, when
$D\neq 0$ and for $T<D^*$ the conductance drops abruptly.~\cite{corna}
A similar $D^*$ was found in an underscreened $S=3/2$ model which displays non-Fermi
liquid behavior for $T < D^*$.~\cite{DinapoliPRL}

Thus, while for temperatures above $D^*$, the model exhibits the features of the isotropic
$S=1$ two-channel model, and the splitting of the triplet states
can be in practice neglected for $T \gg D^*$, at temperatures below $D^*$ one expects
a change of regime and a drop in the conductance. 
This is a general feature of the models for which the spin degeneracy of the impurity is removed,~\cite{zitko} 
as in the underscreened $S=1$ Kondo model with anisotropy.~\cite{corna}

\section{Summary and discussion}
\label{conclusions}

By means of \textit{ab initio} calculations we determined that the minimal amount of O-doping 
needed in a Au chain to push the $5d$-bands towards the Fermi level is 14\%. With
this amount of O atoms, the $5d_{xz,yz}$ orbitals of all the Au atoms present in the chain are conduction
bands, normally below the Fermi level when not doped. Within this minimal concentration of dopants,
we studied the effect of the O atoms on the spin state of a Ni magnetic impurity embedded in the 
O-doped Au chain, as well as the symmetry properties of the unoccupied bands (holes). We showed that
when the O-atoms are fourth neighbors of the Ni impurity, the $S=1$ spin state of the Ni atom comes from
its $3d_{xz,yz}$ empty states. This is a triplet with total angular momentum projection $L_z=0$.
Upon approaching the O atoms towards the Ni impurity, the $3d_{xz,yz}$
begin to be filled, while the $3d_z^2$ starts to be unoccupied, preserving the $S=1$ spin.
Therefore a transition to a Ni 3d$^8$ configuration with two triplets with $L_z= \pm 1$
becomes possible, although correlations stabilize the $L_z=0$ triplet.

After the description of the system and of the first principles calculations, we introduced 
a mixed valence model for the Ni-O-Au chain in the substitutional configuration for Ni,
assuming that the Ni impurity for two holes is in the $L_z=0$ state.
This can be viewed as a generalized Anderson impurity model, which includes not only the charge fluctuations 
(between 3d$^8$ and 3d$^9$ Ni configurations) but also the spin ones. 
By means of a Schrieffer-Wolff transformation we prove that the model can 
be mapped onto a two-channel spin-1 Kondo model. The SU(2) symmetry of the conduction channels of
the O-doped Au chain allows for a full screening of the spin of the Ni impurity. We found that the 
associated Kondo temperature is experimentally accessible and that it is of the order of $T_K \sim400K$.
Therefore, at low enough temperatures as compared with the Kondo one, the system behaves as a Fermi
liquid and a zero-bias anomaly should appear in transport measurements. We also found that even if 
the low energy model corresponds to an anisotropic one, the expected behavior should be similar
to the isotropic case 
due to the fact that the anisotropy constant is found to be smaller than the isotropic Kondo 
temperature.
A small energy scale $D^*$ is dynamically generated, below which the conductivity should
drop. We hope that this work stimulates an accurate many-body calculation of $D^*$ 
which might be compared with the $D^* \approx \exp[-c(T_K/D)^{1/2}]$ behavior obtained 
in similar models.~\cite{DinapoliPRL,corna}

As pointed out in the seminal work by Nozi\`{e}res and Blandin,~\cite{nozieres-blandin} if the 
relation $n=2S$ is satisfied (as it is actually the case), 
then, at temperatures below the characteristic one given by the Kondo
temperature, $T\ll T_K$, the physics of the system is expected to be the same that 
for a Fermi liquid. This is an example of a fully screened higher spin Kondo effect.
Most of the transition metals in bulk conducting materials are described by this kind
of totally screened Kondo phenomena.~\cite{s-1-coleman} However, this is not the 
usual behavior in low dimension and therefore our model represents a possible 
experimental realization of this kind of physics. 
Other low dimensional examples of a two-channel spin-1 Kondo effect can be found in quantum dots (QD).
Specifically, Pustilnik and Glazman~\cite{pustilnik-glazman-1, pustilnik-glazman-2}
studied the possibility of having a spin-1 QD coupled to two screening channels
which fully screen the QD spin. However, they found that even with a small 
difference between the two
antiferromagnetic coupling constants (which corresponds to a real situation), 
the physics of the QD is expected to be dominated by
the underscreened one-channel spin-1 Kondo model, leading to a two-stage Kondo effect.~\cite{coleman-two-stage}
Fortunately, our model based on the O-doped Au chain has the advantage of having 
a protected SU(2) symmetry between the two conducting channels due to the cylindric 
symmetry.

The case with two degenerate triplets in the 3d$^8$ configuration, in which one hole occupies the
3d orbital of $z^2$ symmetry and the other one can be occupied by another hole with
$xz$ or $yz$ symmetry is completely analogous to the case of iron(II) phtalocyanine molecules 
on Au(111).~\cite{mina} In this case, due to the different hybridization of
the Fe orbitals with the substrate (as in our case, the hopping is larger for the 
$z^2$ orbital), a two-stage Kondo effect takes place. First, the spin of the $z^2$ orbital
is screened. It is difficult for us to estimate the Kondo scale for this screening due to 
the uncertainty in the position of the $3d_{z^2}$ state.
The remaining degrees of freedom, the orbital ($xz$ or $yz$) degeneracy and spin 1/2 of the
remaining hole are screened in a rather exotic SU(4) Kondo effect. The effects of the orbital
degeneracy become evident when the molecules are assembled in a lattice.~\cite{tsuka,lobos,joaq}
It would be certainly interesting to find an SU(4) Kondo effect in nanoscopic chains.

\section*{Acknowledgments}

This work was partially supported by PIP No 112-200801-01821, 00273, 112-201201-00069 and 00832 of CONICET, 
and PICT 2010-1060 and 2013-1045 of the ANPCyT, Argentina.

\section{References}
\bibliography{chains.bib}
\end{document}